\documentclass[12pt]{iopart} 
\usepackage{graphicx}
\usepackage[usenames]{color}
\usepackage{setstack}
\usepackage{iopams}

\begin{document}

\title[Transient chaos enforces uncertainty in the British power grid]{Transient chaos enforces uncertainty in the British power grid}

\author{Lukas Halekotte$^1$, Anna Vanselow$^1$ and Ulrike Feudel$^1$}
\address{$^1$ Institute for Chemistry and Biology of the Marine Environment, Carl von Ossietzky University Oldenburg, Oldenburg, Germany}

\ead{lukas.halekotte@uol.de}
\vspace{10pt}
\begin{indented}
\item[]February 2021
\end{indented}

\begin{abstract}
Multistability is a common phenomenon which naturally occurs in complex networks. If coexisting attractors are numerous and their basins of attraction are complexly interwoven, the long-term response to a perturbation can be highly uncertain. We examine the uncertainty in the outcome of perturbations to the synchronous state in a Kuramoto-like representation of the British power grid. Based on local basin landscapes which correspond to single-node perturbations, we demonstrate that the uncertainty shows strong spatial variability. While perturbations at many nodes only allow for a few outcomes, other local landscapes show extreme complexity with more than a hundred basins. Particularly complex domains in the latter can be related to unstable invariant chaotic sets of saddle type. Most importantly, we show that the characteristic dynamics on these chaotic saddles can be associated with certain topological structures of the network. We find that one particular tree-like substructure allows for the chaotic response to perturbations at nodes in the north of Great Britain. The interplay with other peripheral motifs increases the uncertainty in the system response even further.
\end{abstract}
\vspace{2pc}
\noindent{\it Keywords\/}: transient chaos, uncertainty, multistability, chaotic saddle, complex networks, power grid, Kuramoto model with inertia

\vspace{2pc}
\noindent{Prepared using the IOP Publishing \LaTeXe\ preprint class file
\maketitle}

\section{Introduction}

The Kuramoto model with inertia (KM$^+$) has been the subject of tremendous research efforts within the last decade \cite{rodrigues2016kuramoto}. Certainly, one of the main reasons is that it has become common practice to use networks of nonlinear oscillators -- such as the KM$^+$ -- as coarse-scale representations of real power grids \cite{filatrella2008analysis, nishikawa2015comparative, anvari2020introduction}. In addition, the interest in conceptual models of power grids has been fueled by the necessary transformation of todays electrical distribution grids. It is in this context that the KM$^+$ has been applied to tackle some of the major challenges which accompany the decarbonization of power supply, like increasing frequency fluctuations \cite{schmietendorf2017impact, haehne2018footprint, haehne2019propagation}, the loss of inertia within the grid \cite{poolla2017optimal, pagnier2019inertia} or the progressive decentralization of power generation \cite{rohden2012self, rohden2014impact}. 

A main theme in studies modelling power grids as networks of nonlinear oscillators is maintaining a stable operation and thus the avoidance of failures. In particular, the KM$^+$ is suited to unravel the relation between network topology and grid stability. Approaches to this issue are manifold and include various stability concepts based, on the one hand, on different perturbation scenarios such as arbitrary small perturbations \cite{coletta2016linear}, specific \cite{filatrella2008analysis, rohden2012self} or random \cite{menck2014dead} large perturbations or stochastic fluctuations \cite{schafer2017escape, tyloo2019noise} and, on the other hand, on the considered system response, like transient excursion \cite{hellmann2016survivability, tyloo2019key}, asymptotic behavior \cite{menck2014dead} or both \cite{nitzbon2017deciphering} or the severity of cascading failures \cite{rohden2016cascading, schafer2018dynamically}.

\begin{figure}[ht]
\centering
\includegraphics[width=0.85\linewidth]{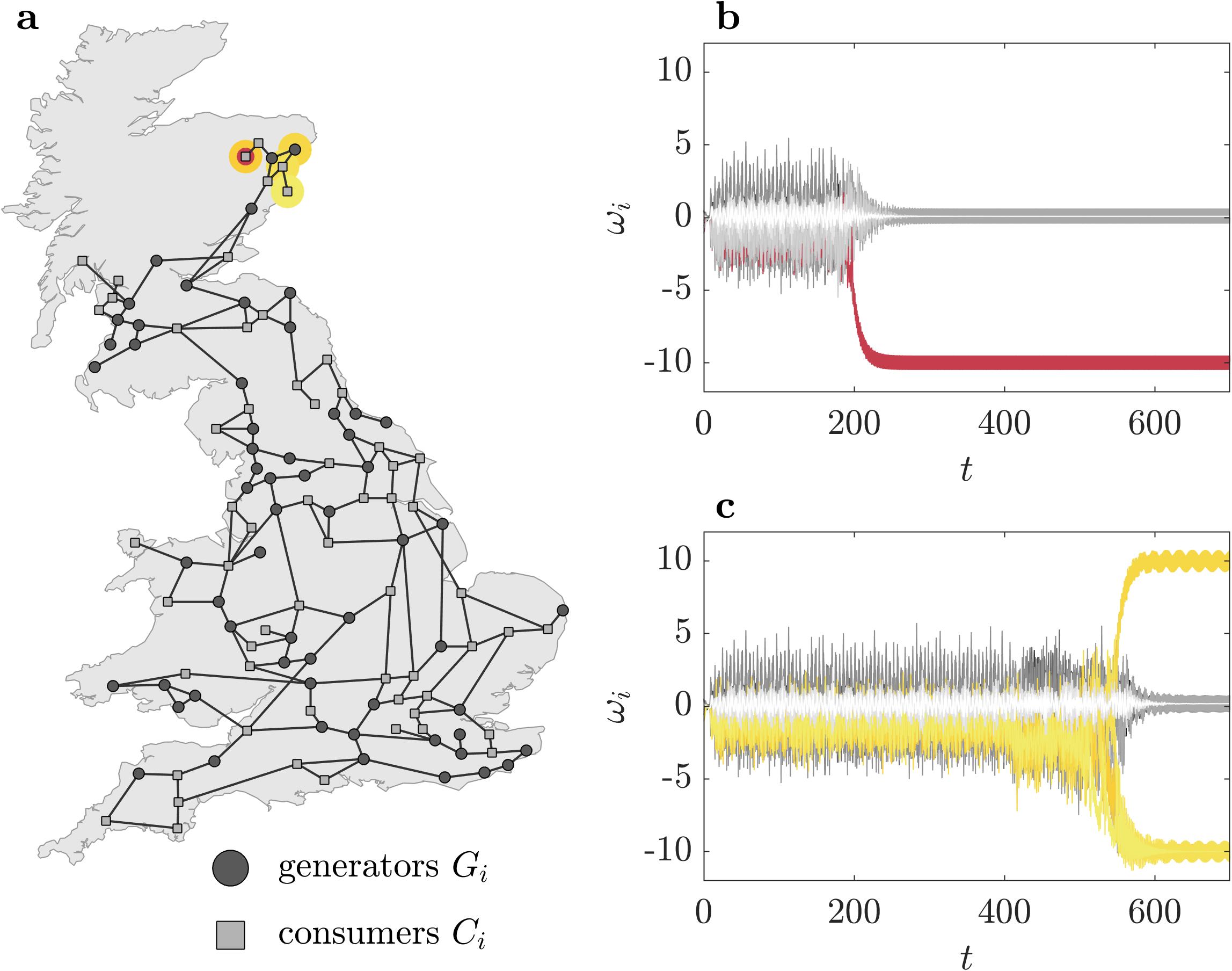}
\caption{(a) Topology of the British power grid \cite{simonsen2008transient}. (b,c) Different outcomes following the 'same' perturbation. The perturbation vector applied in (c) is only $1.0+10^{-7}$ times longer than the vector in (b) and points in the same direction. The nodes which are desynchronized in the end are colored in red and yellow, respectively.}
\label{GB_intro}
\end{figure}

The most fundamental approaches use properties which are inherent in the system to quantify its stability. For instance, if small perturbations are considered, eigenvalues of the operating state can be consulted, whereas properties of its basin of attraction are more suitable if perturbations under consideration are large. The latter explicitly takes into account that power grids -- and thus reasonable parametrizations of the KM$^+$ -- exhibit multistability. In fact, the phase space can be populated by multiple desired synchronized states \cite{coletta2016topologically, manik2017cycle} representing different operating modes as well as several undesired non-synchronized states representing power outages \cite{nitzbon2017deciphering, olmi2014hysteretic, kim2018multistability, hellmann2020network}. In a recent study \cite{halekotte2020minimal}, we contributed to the field of multistability-based stability analyses by determining the minimal fatal shock in a Kuramoto-like representation of the British power grid (figure \ref{GB_intro}(a)). The minimal fatal shock complements the most commonly applied basin stability approach \cite{menck2014dead, kim2018multistability, kim2016building, feld2019large, galindo2020decreased} as it considers another property of the basin landscape (see e.g. \cite{soliman1989integrity, walker2004resilience, mitra2015integrative} for elaborated explanations on this subject): While the basin stability quantifies the stability of the desired state based on the share of volume its basin takes within a predefined frame \cite{menck2013basin, feudel1996map}, the minimal fatal shock uses the shortest distance between the desired state and its basin boundary as a criterion of stability. 

When determining the minimal fatal shock, the only criterion concerning the long-term behavior was that the corresponding perturbation should be fatal. In the framework of the KM$^+$, the attribute fatal simply means that a perturbation induces the desynchronization of at least one oscillator or, in other words, that it pushes the system into any basin of attraction but the one of the fully synchronized state. While we were able to calculate the direction and the magnitude with reasonable precision, the precise outcome of the minimal fatal shock was uncertain. In fact, the slightest variation -- either in the perturbation vector or the integration precision -- could lead to severely different final states, characterized by distinct sets of desynchronized oscillators (figure \ref{GB_intro}(b),(c)). A high sensitivity towards small variations in the perturbation is an indicator for complexly interwoven basins of attraction possessing fractal boundaries \cite{mcdonald1985fractal, feudel2008complex}. This impression appears to be confirmed by the erratic -- seemingly chaotic -- transient dynamics following the fatal shock (figure \ref{GB_intro}(b),(c)). If the dynamics are indeed temporarily chaotic, they showcase the influence of a non-attracting chaotic invariant set -- a so-called chaotic saddle -- embedded within the basin boundary \cite{lai2011transient, grebogi1986metamorphoses, nusse1989procedure}.

Concerning the outcome of specific perturbations, the fractal dimension of basin boundaries is a sufficient determinant of its predictability. If basin boundaries are indeed fractal, small uncertainties in the initial conditions transfer into large uncertainties in the final state which is the classic case of final state sensitivity \cite{grebogi1983final, mcdonald1985fractal}. If however perturbations are only vaguely defined which means that already the initial conditions hold some level of uncertainty, not only the dimension of basin boundaries, but further aspects like the extension of the fractal domains and the number of distinct basins within the -- now larger -- considered region of the basin landscape are decisive \cite{daza2016basin, nieto2020measuring}. In network applications, a typical constraint regards the number of elements, e.g. nodes or edges, which are affected by an initial perturbation \cite{rohden2016cascading, schafer2018dynamically, witthaut2016critical, mitra2017multiple}. For instance, in networks exhibiting multistability, often only the initial conditions of one specific node are varied at a time \cite{menck2014dead, nitzbon2017deciphering, medeiros2018boundaries, dos2020basin}. Huge advantages of such single-node perturbations are that they are usually intuitively interpretable and that the location of the initial impact is well-defined. In this work, we will make use of the latter to examine the relation between the network topology and the uncertainty in the outcome of perturbations. Our studies are thereby based on local basin landscapes corresponding to single-node perturbations within the British power grid.

Although the fractality of basin boundaries \cite{kim2018multistability} and the diversity of non-synchronous states \cite{menck2014dead} have been teased occasionally, most investigations studying the KM$^+$ as a representation of a power grid did not elaborate on the outcome of a harmful perturbation. In fact, due to the focus being on maintaining the synchronous state, it is often only distinguished between safe and unsafe perturbations (e.g. \cite{halekotte2020minimal}). However, this rough classification is systemically justified as the KM$^+$, which neglects voltage dynamics \cite{schmietendorf2014self}, ohmic losses \cite{hellmann2020network} and control mechanisms \cite{schafer2018dynamically}, provides a valid approximation of the synchronization dynamics in real power grids only on short time scales \cite{machowski2020power}. Nevertheless, we believe that an analysis of the complexity and uncertainty in the basin landscape of the KM$^+$ will provide valuable insights, especially concerning the interplay between topology and uncertainty. For instance, the impact of chaotic invariant sets might be significant even on short time scales and might determine the severity of failures within power grids.

The whole work is based on the British power grid \cite{simonsen2008transient} which has been established as a benchmark for power grid analyses \cite{rohden2012self, coletta2016linear, hellmann2016survivability, witthaut2016critical, mitra2017multiple, witthaut2012braess, manik2017network, delabays2017size}. To address the issue of uncertainty within this grid, we proceed as follows: We start by introducing the model equations and the chosen parametrization (section 2). In the following (section 3), we concentrate on cross sections of the basin landscape obtained by perturbing the state variables of single oscillators. Based on these local basin landscapes, we analyze the distribution of uncertainties throughout the grid. We find that uncertainties vary strongly and that some topological features in the northernmost part of the grid can be related to particularly complex basin structures. Therefore, we continue by examining some of the local landscapes in the northernmost area in more detail. In this context, special emphasis is given to the invariant sets of saddle-type found in two cross sections which are responsible for the high complexity in the corresponding landscapes and the high diversity of accessible attractors. Ultimately, we provide a short parameter study and conclude with a discussion of our results (section 4).

\section{The British power grid}

\subsection{Kuramoto-like representation of the British power grid}
We consider the Kuramoto model with inertia (KM$^+$) as a model which captures the desynchronization dynamics within power grids \cite{filatrella2008analysis, rohden2012self}. In this framework, a power grid is described as a network of phase oscillators whose dynamics are given by
\begin{eqnarray} \label{Eq_power}
\nonumber \frac{\mathrm{d} \phi_i}{\mathrm{d} t} &= \omega_i \\
\frac{\mathrm{d} \omega_i}{\mathrm{d} t} &= P_i - \alpha \omega_i + \sum^{N-1}_{j=0} K_{ji} \, \sin(\phi_j-\phi_i) \; ,
\end{eqnarray}
where $\phi_i$ and $\omega_i$ denote the phase and frequency deviation of oscillator $i$ from the grid's rated frequency (hereinafter $\phi_i$ and $\omega_i$ will simply be called phase and frequency). In accordance with the interpretation as a power grid, the parameters $\alpha$ and $P_i$ are the grid's damping constant and the net power input/output of oscillator $i$, respectively. The capacities of the transmission lines and therefore also the topology of the grid is contained in the matrix $K$, with $K_{ji}=K_{ij}>0$ if oscillators $i$ and $j$ are connected and $K_{ij}=0$ otherwise.

Throughout this work, we consider only one network which represents the high-voltage transmission grid of the United Kingdom \cite{simonsen2008transient} (figure \ref{GB_intro}(a)). The grid consists of 120 nodes and 165 transmission lines and the parametrization is chosen in accordance with our former work \cite{halekotte2020minimal}: We assume one half of the oscillators to be generators ($P_i=+P_0$) and one half to be consumers ($P_i=-P_0$). The distribution of generators and consumers within the grid has been drawn randomly. Furthermore, the maximum capacity is the same for all transmission lines and thus either $K_{ij}=K_0$ or $K_{ij}=0$. With $\alpha=0.1$, $P_0=1.0$ and $K_0=5.0$, the model parameters are chosen in order to ensure the coexistence of (at least) one desired synchronous state, with constant phases $\phi^*_i$ and frequencies $\omega^*_i=0$ for all $i \in[0, 119]$, representing stable operation and several undesired non-synchronized states with $\omega_i\neq 0$ for at least one node $i$. We obtain the synchronous state by setting the initial phases and frequencies to $\phi_i=0$ and $\omega_i=0$ for all nodes and integrating the system until it reaches a stable state. The perturbations which are considered in this work are always applied to this stable state and thus will be specified by their phase $\Delta \phi_i$ and frequency difference $\Delta \omega_i = \omega_i$ in the following.

\subsection{Solitary nodes and weak detachment}
Aside from the synchronous state, the KM$^+$ has been shown to hold a variety of different attractors. For instance, the network of oscillators can be split into multiple clusters which are decoupled from each other but internally synchronized \cite{olmi2014hysteretic}. Accordingly, emerging clusters can be characterized by the common time-average of their oscillator's frequencies $\langle \omega_i \rangle$.

Another class of non-synchronous attractors which is expected to be found quiet often -- especially for single-node perturbations -- are solitary states \cite{hellmann2020network} in which one (1-solitary) or more ($n$-solitary) nodes are effectively decoupled from the rest of the grid and swing around their natural frequency $\tilde{\omega}_i$ (here $\tilde{\omega}_i=P_i/\alpha = \pm P_0/\alpha = \pm 10$) while the rest of the network forms a synchronized cluster. In the course of this work, we will refer to corresponding nodes as \textit{solitary detached nodes}. 

Furthermore, Nitzbon \etal \cite{nitzbon2017deciphering} reported another, less common class of attractors in which the detachment of a solitary node is seemingly incomplete or weak.  The weak detachment of a node manifests in a mean frequency lower than its natural frequency and in a higher amplitude of frequency fluctuations. In a way, the \textit{weakly detached node} still feels the impact of the opposing cluster. 

A network of the size of the British grid naturally allows for a large number of coexisting solitary states and cluster states. Moreover, diverse composites which contain multiple clusters and solitary detached nodes are possible. We therefore assume the KM$^+$-representation of the British power grid to be highly multistable.

\subsection{Dead ends and trees}

In this work, topological characteristics which allow for an easy detachment of subparts of the grid are of special interest. For the sake of clarity, we will give a short definition of the most important concepts which we use throughout this work. 

The first concept of relevance is a \textit{tree}. A tree is defined as a connected graph which contains no loops \cite{newman2018networks}. Clearly, the British grid is no tree. However, it involves subgraphs which are tree-shaped \cite{nitzbon2017deciphering} and some which resemble a tree-shaped part.  A tree-shaped part is a subgraph which, if separated from the rest of the network, due to the removal of a single node, fulfills the definition of a tree. The simplest tree-shaped part is a \textit{dead end} which is simply defined as a node with a single link (degree 1). The significance of dead ends in KM$^+$-representations of power grids has already been stressed \cite{menck2014dead, nitzbon2017deciphering}. It should be noted that dead ends and trees are strongly linked to the network theoretical concepts of \textit{articulation points} and \textit{bridges} which denote nodes and edges whose removal would cut a network into multiple subgraphs. 


\section{Uncertainty in the British power grid}

\subsection{The coloring of nodes \label{sec_coloring}}
The scope of this work can be roughly summarized as examining the relation between the location of a perturbation and the uncertainty of its outcome. By the outcome of a perturbation we mean the asymptotic long-term behavior or attractor which the system approaches after being perturbed. In order to ensure that their location is well-defined, we focus on perturbations which, in each case, only affect the phase $\phi_i$ and frequency $\omega_i$ of a single node $i$. Since we additionally assume that the system is situated in the fully synchronized state prior to a disturbance, all perturbations are basically drawn from two-dimensional cross sections of the high-dimensional phase space. In the following, we will use properties of these cross sections to assess the distribution of uncertainty in the KM$^+$-representation of the British power grid.

We start our analysis by explicitly inspecting each single-node cross section in the British grid. To this end, we assign a set of $750\times750$ initial conditions to each node, wherein phase and frequency deviations at node $i$ are equally distributed in a frame with $\Delta \phi_i \in [-\pi, \, \pi]$ and $\Delta \omega_i =\omega_i \in [-15, \, 15]$, while $\Delta \phi_j=0$ and $\omega_j=\omega^*_j = 0 \; \forall \; j \neq i$. Using numerical integration \cite{ansmann2018efficiently}, we then determine the attractors to which the initial conditions converge. Attractors are thereby differentiated on the basis of the temporal mean and the fluctuation magnitude of nodal frequencies \cite{nitzbon2017deciphering, olmi2014hysteretic}: We assume that two trajectories belong to the same attractor if, after a sufficiently long transient (maximum integration time $t_{max}=5\cdot 10^5$), the differences between their temporal mean, minimum and maximum frequency are smaller than predefined thresholds ($0.2$, $0.7$, $0.7$) for each $\omega_j$ with $j \in [0,\,119]$. The corresponding thresholds were obtained empirically. Ultimately, by relating each initial condition in every cross section to an attractor, we obtain a \textit{local basin landscape} for each node. With respect to Daza \etal \cite{daza2016basin} and to the usual presentation of basins, we also refer to the relation between initial condition and attractor as the \textit{color} of that initial condition. 

\begin{figure}[ht]
\centering
\includegraphics[width=1.0\linewidth]{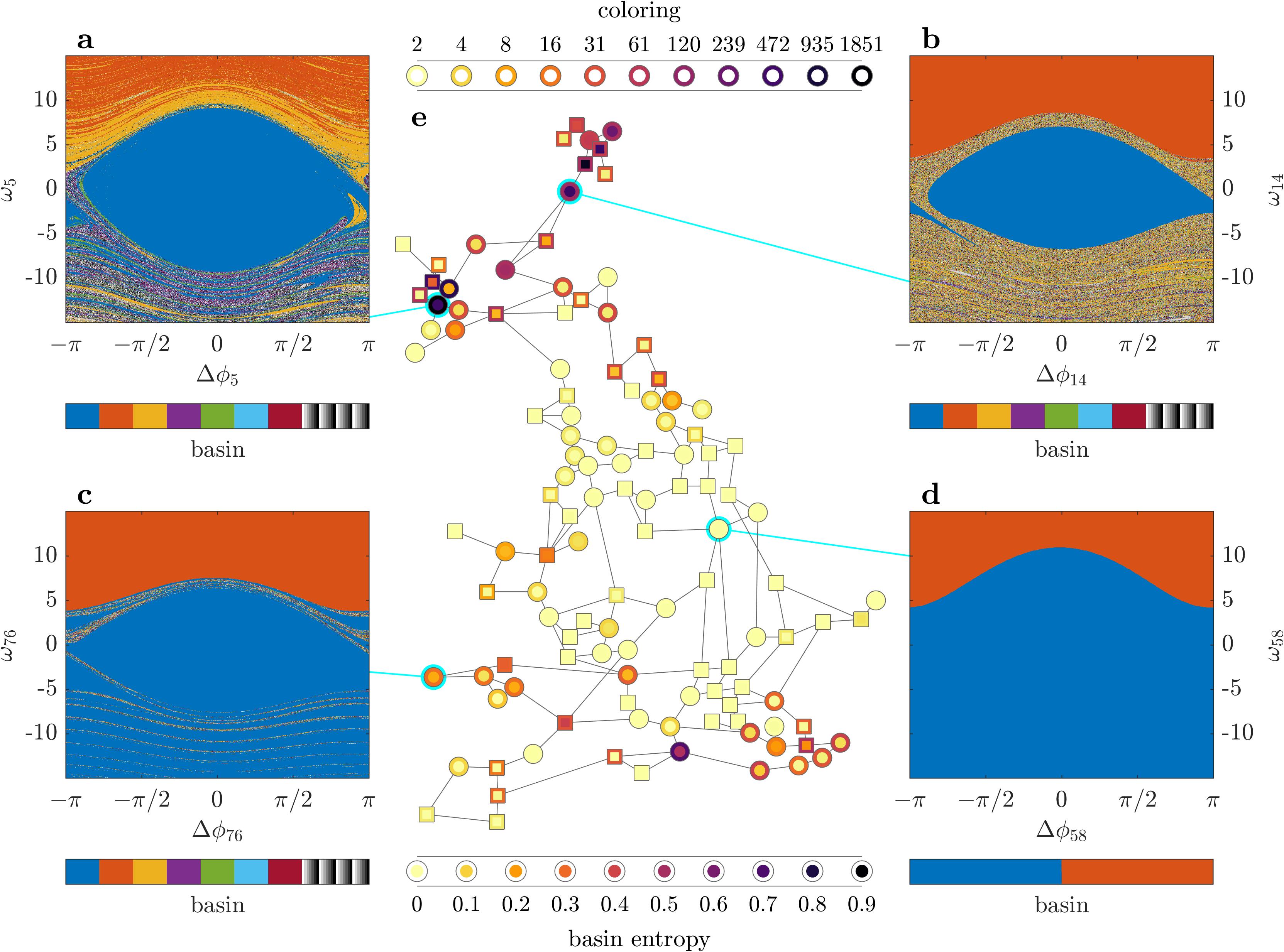}
\caption{Uncertainty in the British power grid. (a-d) Local basin landscapes of four exemplary nodes: (a) N5, (b) N14, (c) N76, (d) N58. Each color depicts a different basin of attraction. While the darker blue always corresponds to the basin of the fully synchronized state, the remaining basins are sorted according to their relative size in the landscape. The seven most common basins receive a specific color while additional basins are colored in recurring shades of gray. (Middle column) Basin entropy (inner color) and coloring (edge color) of all nodes. Circles portray generators and squares consumers.}
\label{GB_coloring}
\end{figure}

The visual inspection of four exemplary local basin landscapes (figure \ref{GB_coloring}(a)-(d)), each corresponding to a generator in the network, shows that their coarse structure bears some similarity, but that their complexity differs immensely. We find that each local landscape includes two, more or less dominant, characteristic basins, corresponding to the fully synchronized state (blue) and a non-synchronous state in which the perturbed node itself is solitary detached (orange). However, while the smooth boundary between these two basins represents the only boundary in the local landscape of N58 (figure \ref{GB_coloring}(d)), the landscapes of N5, N14 and N76 (figure \ref{GB_coloring}(a)-(c)) contain domains of multiple, highly intertwined basins. The fine structure and extent of these domains again varies strongly ranging from rather thin stripes (figure \ref{GB_coloring}(c)) to extended areas (figure \ref{GB_coloring}(b)).

Differences in the structure and complexity of local basin landscapes imply differences in the uncertainty of the outcome of single-node perturbations. In the following, we will address the distribution of uncertainties within the British grid based on two attributes which capture appropriate characteristics of the local basin landscapes: the \textit{coloring} and the \textit{basin entropy}. As the coloring of a node we simply denote the number of colors found in the corresponding landscape. Since every color represents an attractor, the coloring is a seemingly natural indicator of the number of possible outcomes due to perturbations at the specific node. It should, however, be noted that -- due to the finiteness of the set of initial conditions, the uneven distribution of complex domains and the potential length of some transients -- the coloring itself holds a certain degree of uncertainty. In fact, depending on the distribution of basins within a local landscape, the real number of accessible attractors can be severely underestimated (see \ref{append_a}). Nevertheless, we assume that the coloring is instructive since it allows us to gain some insight into the diversity of accessible attractors.

The coloring is, however, not suitable as an exclusive measure of uncertainty since it neglects essential aspects like the frequency of colors or the mingling of corresponding basins. Suitable measures which could complement the coloring in this regard are the basin stability \cite{menck2013basin} and the uncertainty exponent \cite{grebogi1983final}, both of which have been applied in similar contexts to capture one of these aspects \cite{medeiros2018boundaries, dos2020basin}. We choose neither of the two but follow a more holistic approach which has been introduced by Daza \etal \cite{daza2016basin}. Their measure -- called \textit{basin entropy} -- has been shown to be sensitive to the fractality and size of basin boundaries as well as to the number of colors. It thus incorporates multiple aspects of uncertainty within a single index. In order to obtain a nodewise uncertainty index, we apply a single node-version of the basin entropy which is based on the local basin landscapes. The basin entropy $S_b$ is defined as
\begin{equation} \label{eq_entropy}
S_b = \frac{1}{N_b} \sum^{N_b}_{k=1} \left(- \sum^{m_k}_{l=1} p_{kl} \log p_{kl} \right) \; .
\end{equation} The procedure to obtain $S_b$ can be outlined as follows: Divide a region in phase space -- in our case, one of the cross sections -- into $N_b$ boxes, calculate the Gibbs entropy for each box $k$ (term in brackets in (\ref{eq_entropy})) and average over all boxes. The calculation of the Gibbs entropy is thereby based on the distribution of colors in the corresponding box. In this context, $m_k$ denotes the number of colors found in box $k$ and $p_{kl}$ the probability of an arbitrary initial condition in this box to hold the specific color $l$ (approximated by the number of initial conditions with color $l$ divided by the total number of initial conditions in box $k$). We calculate the basin entropy for each node (figure \ref{GB_coloring}(e)) based on the set of $750\times750$ initial conditions which we divide into $150\times150$ equally sized boxes, each holding 25 initial conditions. 

The distribution of the coloring and basin entropy over the British grid shows that the uncertainty in the outcome of a perturbation exhibits strong variations, both globally and locally (figure \ref{GB_coloring}(e)). We find that large parts of the grid -- especially in the center -- possess minimal uncertainties including a coloring of $2$ (light yellow coloring). On the contrary, we find several peripheral areas -- e.g. in the southeast, in the west and in the north in particular -- in which the outcomes of localized perturbations are highly uncertain (darker colors). It is notable that the proximity of nodes with extraordinary high values of basin entropy and coloring always includes other nodes whose basins exhibit high complexities as well. An observation which might suggest a relation between uncertainty and specific larger topological structures. Nevertheless, we also find that within areas of high uncertainty, basin entropy and coloring vary strongly among nodes. For instance, nodes which are easily detached -- like dead ends -- exhibit lower uncertainties than adjacent nodes which show relatively high values of basin entropy and coloring.

\subsection{Scotland in the spotlight}
\begin{figure}
\centering
\includegraphics[width=1.0\linewidth]{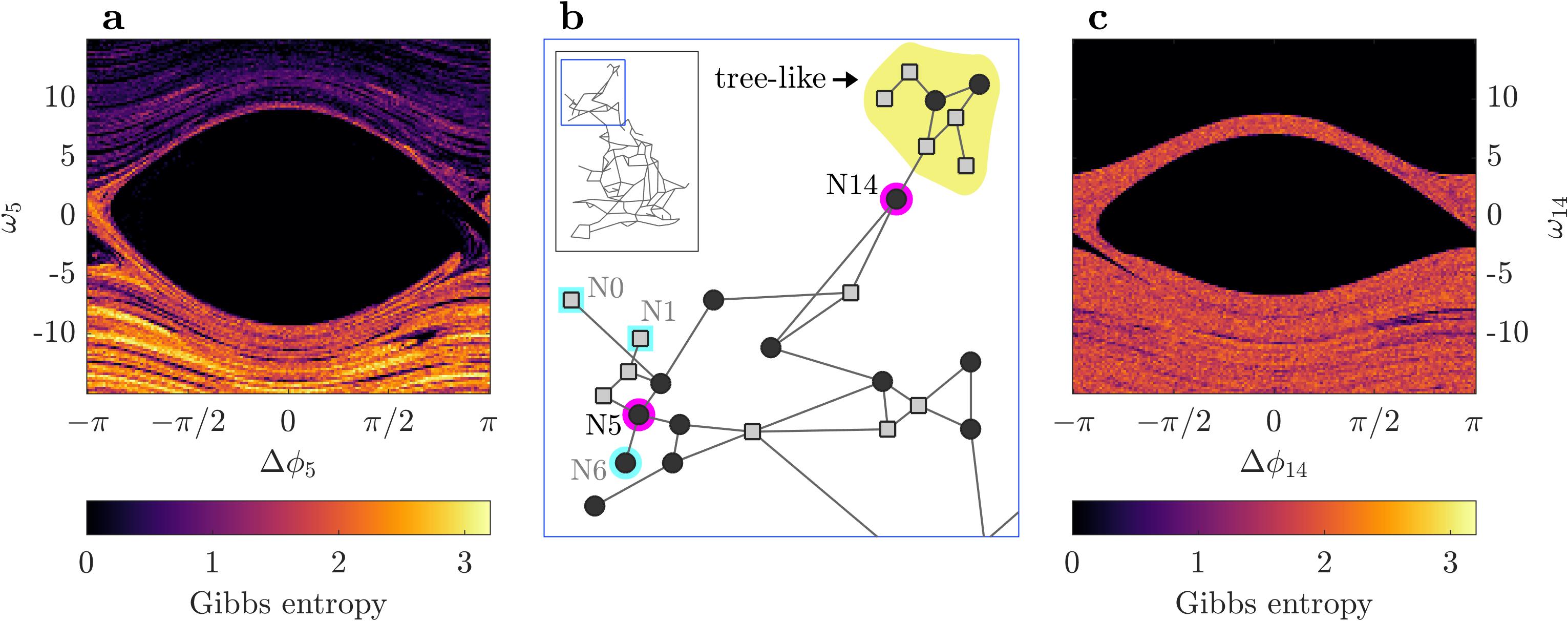}
\caption{Spatial resolution of the Gibbs entropy -- term in brackets in (\ref{eq_entropy}) -- for the nodes N5 (a) and N14 (c). Minimum and maximum Gibbs entropies are $0$ (black) and $-log(1/25) \approx 3.22$ (yellow), respectively. Minimum/maximum: All 25 initial conditions in one box approach the same/different attractor(s). (b) Enlarged display of the northern part of the British grid. The nodes N5 (a) and N14 (c) are framed in magenta. The yellow shaded region highlights the tree-like substructure. As a reference for upcoming results, the dead ends N0, N1 and N6 are marked in cyan. Circles represent generators and squares consumers.}
\label{GB_gibbs}
\end{figure}

Particularly high indices of uncertainty are found in the northernmost or 'Scottish' part of the grid (figure \ref{GB_gibbs}(b) and figure \ref{GB_coloring}(e)). Especially in the northeast, the high basin entropies of several adjacent nodes are striking. Importantly, this region of pronounced uncertainty coincides with an exceptional topological structure which resembles a dead end in the sense that it also holds a single connection (bridge originating at N14; figure \ref{GB_gibbs}(b)) to the rest of the grid. We refer to the corresponding structure as \textit{tree-like} (yellow shaded region in figure \ref{GB_gibbs}(b)) since it contains multiple bridges and articulation points which are characteristic for \textit{trees} but one more edge than an actual tree. 

Despite the accumulation of nodes with high uncertainty around the tree-like, the highest coloring (1851) is found elsewhere, rather distant at N5 (still in Scotland!). In fact, not only the most but also the second (696) and third (458) most colorful node is located in this area. Compared to the highest coloring found at the tree-like (coloring of 101 at N14), all three allow for a massive number of possible outcomes. Topologically, the close proximity of three associated dead ends in this region is striking: Each of the three most colorful nodes is linked to a dead end (N0, N1 and N6 ; figure \ref{GB_gibbs}(b)).

In order to uncover the origin of the high uncertainties in these two distinct regions, we will examine two exemplary cross sections in more detail. We choose the node N5 as it exhibits by far the highest coloring as well as the highest basin entropy apart from the tree-like and its single neighbor N14. As a representative of the tree-like, we consider N14 as it represents, in a way, the topological equivalent to N5: Both are articulation points which link a peripheral motif -- dead end and tree-like, respectively -- to the rest of the grid.

The basin boundaries in the complex domains of both local basin landscapes are indeed fractal (uncertainty exponent $\gamma \approx 0.02$ for N5, $\gamma \approx 0.03$ for N14; see \ref{append_b}). Furthermore, according to the basin entropy (\ref{eq_entropy}), the uncertainty of the two nodes is pretty much the same (N5: $S_b\approx 0.732$, N14: $S_b\approx0.736$). Nevertheless, the landscapes differ significantly: While the complexity at N14 (figure \ref{GB_coloring}(b)) stems from one rather consistently mixed fractal domain, the local landscape of N5 (figure \ref{GB_coloring}(a)) contains multiple fractal domains which differ with regard to the color composition and the complexity of the interwoven basins. This impression is backed up by the spatial resolution of the Gibbs entropy (figure \ref{GB_gibbs}(a),(c)). Particularly noteworthy are the strong differences in the local landscape of N5, with the highest entropies in boxes corresponding to negative frequency perturbations. Some boxes at N5  even reach the maximum Gibbs entropy of $-log(1/25)$ which means that from a set of $5\times5$ initial conditions, each one converges towards a different attractor.

\subsection{What's transient chaos got to do with it}
\begin{figure}[ht]
\centering
\includegraphics[width=0.9\linewidth]{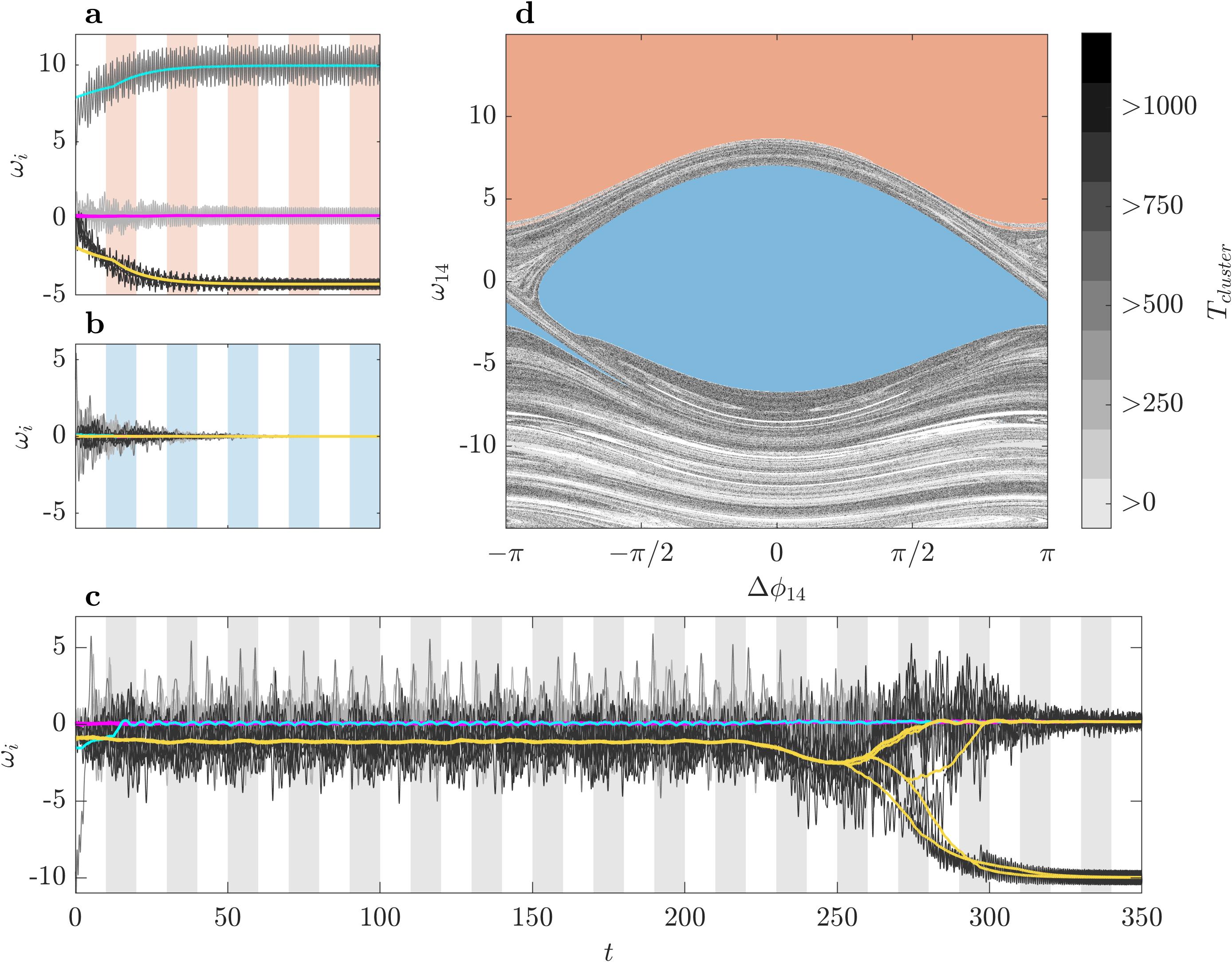}
\caption{(d) Division of the local basin landscape of N14 into three regions -- blue, red and grayscale -- owing to three distinct transient behaviors. The different shades of gray denote the lifetime of the tree-like cluster. (a-c) Exemplary trajectories, each corresponding to one of the three regions (corresponding colored stripes). Thin gray lines depict the actual frequencies and thick colored lines the moving average of the frequencies of N14 (cyan), the nodes in the tree-like (yellow) and in the rest of Scotland (magenta).}
\label{GB_saddleN14}
\end{figure}

In the following, we expand our analysis on the two exemplary local basin landscapes, starting with the node which shows the lower coloring, N14. Despite its high complexity, the local basin landscape of N14 can be divided into three distinct regions. The first two correspond to the basins of the two most commonly approached attractors, characteristic to any landscape: The fully synchronized state (blue region in figures \ref{GB_coloring}(b) and \ref{GB_saddleN14}(d)) and the solitary state in which the perturbed node itself is solitary detached (orange region in figures \ref{GB_coloring}(b) and \ref{GB_saddleN14}(d)). Apart from these two basins, the landscape is filled by a mixed region which is dominated by basin boundaries rather than expanded basins. In this sense, it resembles a riddled structure. Due to the relatively consistent mixing of basins (figure \ref{GB_gibbs}(c)), this area can be denoted as the third coherent region. It should be noted that it is this region which accounts for the high basin entropy (figure \ref{GB_gibbs}(c)) as well as the vast majority of accessible attractors.

Differences between the three regions (blue, orange, mixed) become apparent especially in the transient dynamics. While the convergence of initial conditions in the blue and orange region is straightforward and usually fast (figure \ref{GB_saddleN14}(a),(b)), trajectories initiated in the mixed region often pass rather long transient stages before settling on one of several accessible attractors (figure \ref{GB_saddleN14}(c)). Importantly, most of these transients show a characteristic behavior which involves strong, erratic frequency fluctuations. Calculating the moving average of each oscillator's frequency, we find that a certain topological feature underlies this behavior. Similar to the dominant solitary state (figure \ref{GB_saddleN14}(a)), the seven nodes in the tree-like form a cluster (figure \ref{GB_saddleN14}(c)). However, in contrast to the cluster whose formation is forced by the solitary detachment of N14, this tree-like cluster is seemingly weakly detached from the rest of the grid. This weak detachment shows in the comparatively low magnitude of the mean frequencies of the involved nodes which do not reach the mean of their natural frequencies (compare figures \ref{GB_saddleN14}(c) and \ref{GB_saddleN14}(a)). 

At this point, two essential questions concerning the transient behavior remain unanswered. The first is how common this behavior is and thus how strongly it determines the system's response within the mixed region. We approach this by testing for a set of random initial conditions within the cross section of N14 whether the weakly detached cluster is formed during the initial phase of the transient and, if so, how long it persists. It shows that the set of initial conditions which approach this stage basically coincides with the mixed region (compare figure \ref{GB_saddleN14}(d) and \ref{GB_coloring}(b)): The formation of the weakly detached tree-like cluster is indeed characteristic for the mixed region. The second question regards the irregularity of oscillations which might display the impact of an unstable chaotic invariant set. Based on a pseudo-trajectory (see figure \ref{append_pic_c}(a)) -- which we obtained using the stagger-and-step method \cite{sweet2001stagger} -- we verified that the dynamics are indeed chaotic (see \ref{append_c}). 

In conclusion, the division of the basin landscape of N14 into three dominant regions which was primarily based on our visual impression is backed up by three distinctive qualitative behaviors following a perturbation (figure \ref{GB_saddleN14}(d)): (Blue) The system follows an oscillatory path to its desired state (figure \ref{GB_saddleN14}(b)). (Orange) Node N14 is desynchronized fast due to a strong acceleration and thereby splits the grid into two completely detached clusters (figure \ref{GB_saddleN14}(a)). (Mixed/Gray) The system approaches a chaotic saddle, stays in its vicinity for a while and finally settles on one of multiple accessible attractors (figure \ref{GB_saddleN14}(c)).

\subsection{More chaos at N5}

We have seen that the complex basin landscape at N14 is due to transients which temporary follow an unstable chaotic invariant set. As this chaotic saddle is visibly associated with the tree-like, the interrelationship between network topology and basin complexity becomes apparent. However, the origin of the extraordinary high uncertainty at the rather distant node N5 is still unclear (figure \ref{GB_coloring}(a),(e)). 

\begin{figure}[ht]
\centering
\includegraphics[width=1.0\linewidth]{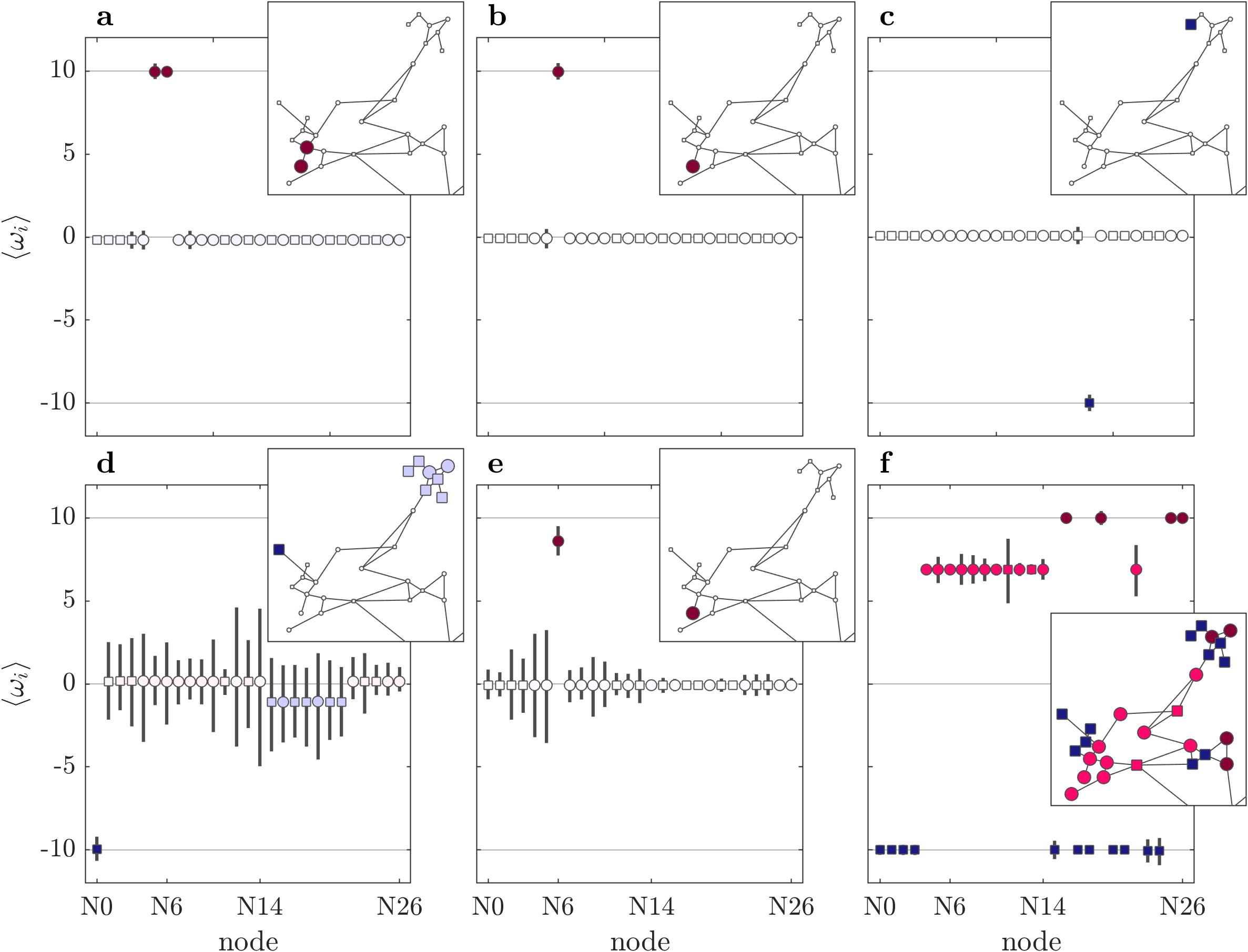}
\caption{Six exemplary attractors approached after a perturbation at node N5. Shown are the mean frequencies (circles and squares) and the range of oscillations bounded by the minimum and maximum frequency value (vertical dark gray lines) for the upper 27 oscillators within the British grid. The color saturation of the markers depict the mean frequencies (blue - negative, red - positive). (a-c) The three most common attractors -- besides the fully synchronized state ($\omega_i=0 \; \forall \; i$) -- which are approached by 93529 (a), 75314 (b) and 23640 (c) initial conditions. (d-f) Less common attractors which are approached by 963 (d), 142 (e) and 5 (f) initial conditions.}
\label{GB_examples}
\end{figure}

To obtain a first hint at the source of its colorfulness, we take a look at some of the attractors which are accessible due to perturbations at N5. The three most common attractors which are the fully synchronized state (not shown), a 2-solitary state in which N5 and its dead end neighbor N6 are detached (figure \ref{GB_examples}(a)) and a 1-solitary state in which only N6 is detached (figure \ref{GB_examples}(b)) are not surprising. Furthermore, finding the topological counterpart of the latter in which N6 is only weakly detached is hardly surprising, although it is comparatively rare (figure \ref{GB_examples}(e)). 

Aside from the few expectable outcomes, the response to perturbations at N5 can be exceptionally complex which shows, for instance, in severe outcomes involving the desynchronization of large parts of the grid (figure \ref{GB_examples}(f)). However, also the accessibility of some of the rather common outcomes is surprising. An illustrative example is a 1-solitary state whose detached node lies within the tree-like and thus far from the originally perturbed node (figure \ref{GB_examples}(c)). Another example resembles the chaotic saddle in the sense that it involves the same weakly detached cluster but one additionally detached dead end, N0 (figure \ref{GB_examples}(d)). Interestingly, both attractors are related to the tree-like which might indicate its impact on the complexity of the local basin landscape of N5.

\begin{figure}[ht]
\centering
\includegraphics[width=0.95\linewidth]{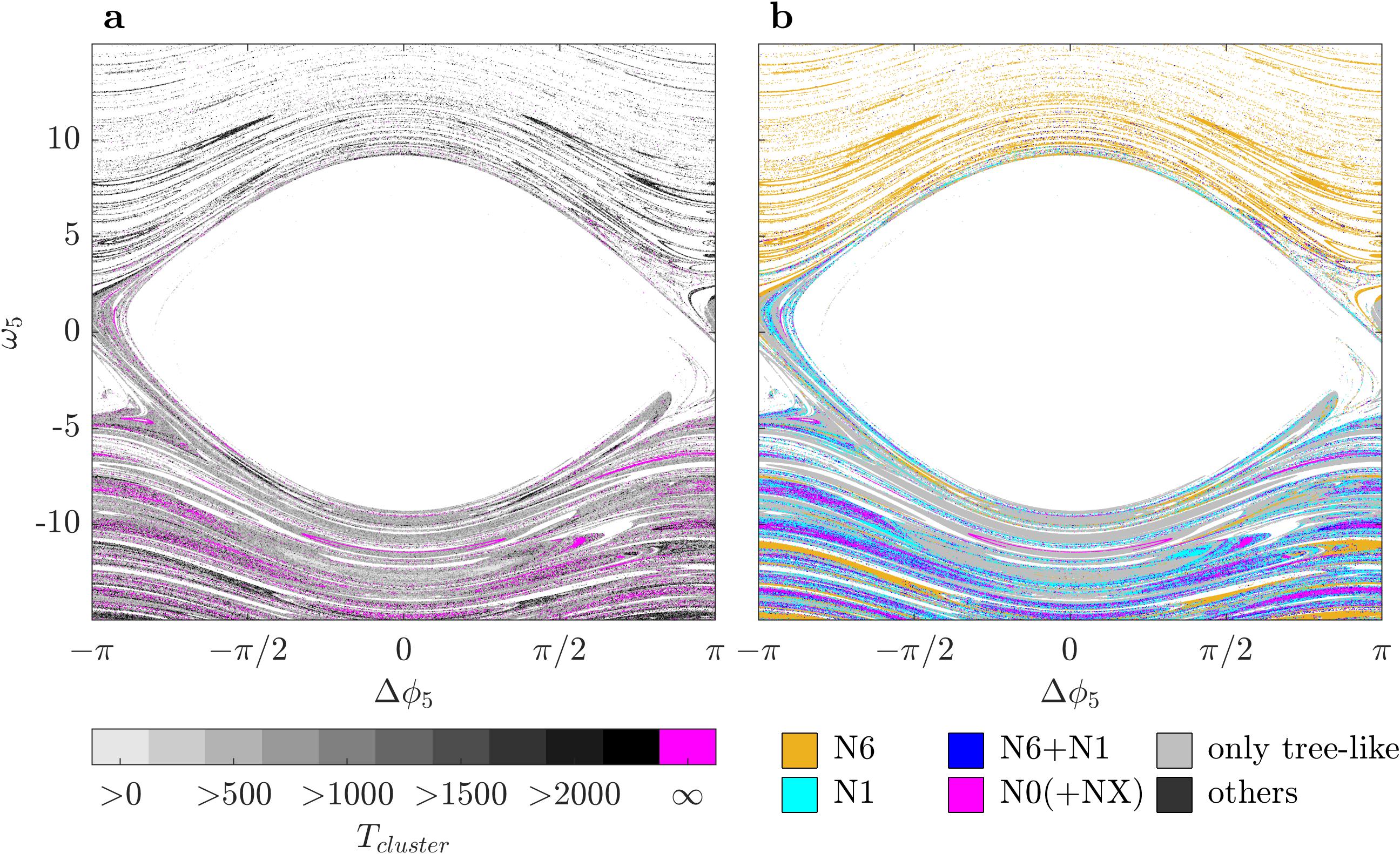}
\caption{Chaotic transients originating from the local basin landscape of N5. (a) The grayscale displays the lifetime of the tree-like cluster. Magenta colored initial conditions approach an attractor in which the cluster persists. The cluster is not formed for initial conditions situated in white areas. (b) Depiction of the nodes (brownish yellow, cyan, dark blue, black) which are detached during the existence of the cluster. Gray points mark the additional desynchronization of no other nodes and magenta points the additional detachment of N0 and any combination of further nodes.}
\label{GB_saddleN5}
\end{figure}

In order to test this, we again take a look at the transient dynamics. We find that trajectories initiated within the mixed areas of the cross section of N5 exhibit a transient stage in which the weakly detached tree-like cluster forms. In accordance with our previous approach, we check for a set of random initial conditions in the cross section of N5 whether the cluster is formed and, if so, how long it persists (figure \ref{GB_saddleN5}(a)). It shows that the mixed regions of the local landscape of N5 are determined by dynamics associated with the weakly detached tree-like cluster as well. However, in comparison to N14 (figure \ref{GB_saddleN14}(d)), we find trajectories in which the tree-like cluster persists much longer, even becoming actually persistent (figure \ref{GB_saddleN5}(a)), at least for the maximum of our integration time.

These long lifetimes come as no surprise since we have already noticed that perturbations at N5 can lead to an attractor which involves the weakly detached tree-like cluster (figure \ref{GB_examples}(d)). The obvious difference between the chaotic transient approached for perturbations at N14 and this attractor is that the node N0 is solitary detached in the latter (figure \ref{GB_examples}(d)). Furthermore, we find multiple additional transient dynamics which involve the weakly detached tree-like cluster but differ with regard to the nodes which are additionally detached, especially N1 and N6. Each combination of additionally detached nodes accounts for a particular chaotic saddle (or attractor) which ultimately explains the extraordinary coloring of N5 -- each saddle paves the way to a distinct set of attractors. A spatial depiction of the temporarily approached saddles shows that in certain regions of the cross section, certain saddles dominate (figure \ref{GB_saddleN5}(b)). In some regions, even the structure of the 'saddle landscape' itself is complex.

\subsection{Uncertain uncertainty}
We have seen that the existence and accessibility of chaotic saddles is a decisive factor determining the uncertainty in the outcome of single-node perturbations. However, so far, our examinations have been entirely based on one particular parametrization of the British grid. In the following, we aim at a brief insight into the variability of the uncertainty depending on varying parameters. As an exemplary parameter we choose $K_0$ which determines the maximum capacity of all transmission lines. Since we need the fully synchronized state as a reference for the positioning of our cross sections, we vary $K_0$ within a range which ensures the existence and stability of this state. For each value of $K_0$, we then calculate the coloring and basin entropy in accordance with our earlier explanations. For the sake of simplicity, we concentrate on one particular node, N14.

\begin{figure}
\centering
\includegraphics[width=1.0\linewidth]{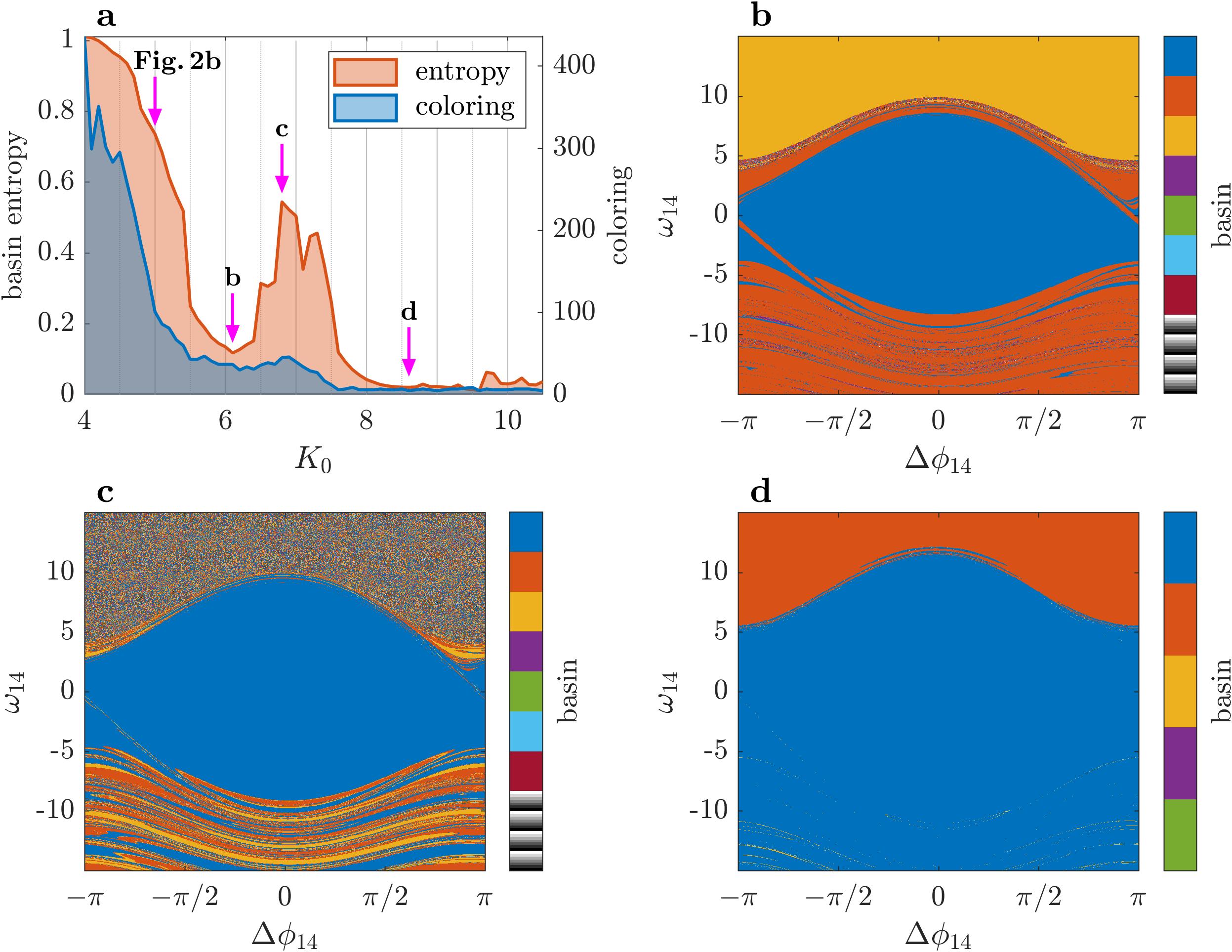}
\caption{Uncertainty in the local basin landscape of N14 depending on the transmission capacity $K_0$. (a) Basin entropy and coloring. Magenta arrows refer to the local basin landscapes shown in figure \ref{GB_coloring}(b) and in the subfigures (b), (c), (d). (b)-(d) Local basin landscapes for $K_0 = 6.1$ (b), $K_0 = 6.8$ (c) and $K_0 = 8.6$ (d). Each color depicts a different basin of attraction. The darker blue corresponds to the basin of the fully synchronized state. The remaining basins are sorted according to their frequency. Only the seven most common basins receive a specific color while the others are colored in recurring shades of gray.}
\label{GB_parameter}
\end{figure}

The uncertainty in the outcome of perturbations is particularly high for a rather wide range of low transmission capacities which coincide with a less voluminous basin of the fully synchronized state (figure \ref{GB_parameter}(a)). In this respect, the decrease in uncertainty with increasing $K_0$ is non-surprising as the basin of the fully synchronized state takes increasing amounts of the tested cross section while, at the same time, the number of accessible attractors diminishes. However, in contrast to the coloring, the progression of the basin entropy is not at all monotonous but does indeed possess pronounced local minima and maxima (especially b and c in figure \ref{GB_parameter}(a)).

The resulting topography -- valley: minima at b in figure \ref{GB_parameter}, peak: maxima at c in figure \ref{GB_parameter} -- emphasizes once again that particular unstable invariant sets determine the complexity of the local basin landscape. To recognize this, a closer look at the dominant regions of the local landscape for some exemplary values of $K_0$ is worthwhile (figure \ref{GB_coloring}(b) and figure \ref{GB_parameter}(b)-(d)): For $K_0=5.0$, the local landscape is populated by three dominant regions of which one can be characterized on account of the transient dynamics which show the impact of a chaotic saddle. As $K_0$ is increased, this chaotic invariant set becomes attractive. Accordingly, the formerly mixed region of initial conditions which was determined by the attraction and subsequent repulsion of the saddle does now approach this attractor (orange domain in figure \ref{GB_parameter}(b)). Since now the local landscape is mainly filled by three basins, the basin entropy exhibits comparatively low values (b in figure \ref{GB_parameter}(a)). The subsequent increase in the basin entropy is then established due to the solitary state losing its stability and thereby making room for a 'new' mixed region (figure \ref{GB_parameter}(c)). Interestingly, at this point, the two attractors whose basins populate most of the local landscape -- aside from the fully synchronized state -- both include the weakly detached tree-like cluster and only differ slightly in the mean frequency within the cluster. Subsequently, the 'new' mixed region is again taken over by the basin of a solitary state in which N14 is detached. At the same time the expansion of other basins diminishes which ultimately leads to basin entropies at constantly low levels.

\section{Discussion}
Inspired by findings in an earlier work \cite{halekotte2020minimal} (figure \ref{GB_intro}), we examined the uncertainty in the outcome of localized perturbations within a Kuramoto-like representation of the British power grid. Using local basin landscapes corresponding to perturbations at single nodes, we demonstrated that the basin complexity and thus the uncertainty in the outcome of a perturbation is highly variable with low uncertainty predominantly in the core of the network and high uncertainty concentrated close to peripheral regions (figure \ref{GB_coloring}). Particularly complex domains in the basin landscape which resemble riddled structures can be related to unstable invariant chaotic sets which are accessible by initial conditions within these areas (figure \ref{GB_saddleN14} and figure \ref{GB_saddleN5}). However, most importantly, we showed that the chaotic dynamics on the saddles are related to certain topological structures within the network. We found that one particular peripheral motif -- the tree-like (figure \ref{GB_gibbs}(b)) -- allows for the chaotic response to perturbations at nodes in the north of Great Britain. Furthermore, this response can be affected by the interplay with other peripheral motifs -- dead ends in the northwest -- which increases the potential for a complex system response even further (figure \ref{GB_saddleN5}).

At this point, we like to emphasize that this work gives only a first insight into the complexity which is inherent to the presented model. In fact, we understand our work also as an expression of the amazement which we felt when studying the dynamics within what is only a single realization of the British grid. To stress this impression, we name here a few observations which we have only briefly touched. The first is the stabilization of chaotic dynamics. We have seen that, in line with other studies on the role of chaotic saddles in synchronization dynamics \cite{medeiros2018boundaries, medeiros2019state}, an increase of the transmission capacity stabilizes the formerly chaotic saddle. The same effect can be obtained by the additional desynchronization of the dead end N0 but, surprisingly, not by the nearby dead end N1 (figure \ref{GB_saddleN5}). Importantly, the stabilization of the chaotic dynamics goes along with a loss of characteristic time intervals in which the system coherence drops (see \ref{append_c}). As these drops represent escape paths from the unstable chaotic set, an understanding of their demise will most likely be insightful but is beyond the scope of this study. 

Another aspect is the diversity of coexisting attractors. Although we have discussed the number of accessible attractors (coloring) and presented some examples, we did not specifically consider the process of their emergence (bifurcations) or the conditions of their existence. Neither did we consider the full range of different attractor types. Furthermore, it is important to notice that the coloring, which we introduced as a qualitative measure, at times largely underestimates the real number of accessible attractors, especially for the most complex local basin landscapes (see \ref{append_a}). The seemingly unlimited number of accessible attractors in some landscapes is probably caused by connections between different unstable invariant sets or possible mergings of chaotic saddles \cite{kraut2002multistability}. An indication for this relationship is found at node N5 where not only the different basins but also areas of initial conditions approaching different chaotic saddles are heavily intertwined. Importantly, the existence of paths which pass multiple unstable invariant sets are a likely cause of extremely severe desynchronization events due to rather insignificant disturbances. Although these outcomes are rare, we emphasize that the existence of chaotic saddles enables relatively small and locally restricted perturbations to cause extreme desychronization events, as exemplified in figure \ref{GB_examples}(f).

To what extent our findings apply to real world electrical distribution grids is unclear since the Kuramoto model with inertia is only a valid representation on very short time scales \cite{machowski2020power}. Accordingly, in order to examine the uncertainty in a more realistic setting, more sophisticated model equations had to be applied \cite{schmietendorf2014self, weckesser2013impact, auer2016impact}. Nevertheless, it is already instructive to notice that this basic approximation to a power grid, which captures crucial aspects of its synchronization dynamics, is capable of exhibiting chaotic behavior. Concerning the control, the presence of chaos would in fact be crucial since the presence of chaotic saddles requires very different control strategies \cite{lilienkamp2020terminating, capeans2017partially, dhamala1999controlling}. 

Furthermore, due to the rather elementary treatment of synchronization, our findings might as well be meaningful in other applications dealing with synchronization phenomena on networks (e.g. \cite{arenas2008synchronization, pikovsky2015dynamics}). This is especially true since, in this work, the emergence of chaos is related to topological features of the network -- in contrast to studies in which chaotic dynamics are inherent to the single units of a network \cite{medeiros2018boundaries}. In this sense, the special role of peripheral structures like dead ends or the tree-like is stressed even further since they might not only affect the stability against perturbations \cite{menck2014dead, halekotte2020minimal} but also the uncertainty in and the severity of the long-term system response.

\ack
The simulations were performed at the HPC Cluster CARL, located at the University of Oldenburg (Germany) and funded by the DFG through its Major Research Instrumentation Programme (INST 184/157-1 FUGG) and the Ministry of Science and Culture (MWK) of the Lower Saxony State.

\appendix

\section{Some comments on the coloring \label{append_a}}
\begin{figure}[ht]
\centering
\includegraphics[width=0.7\linewidth]{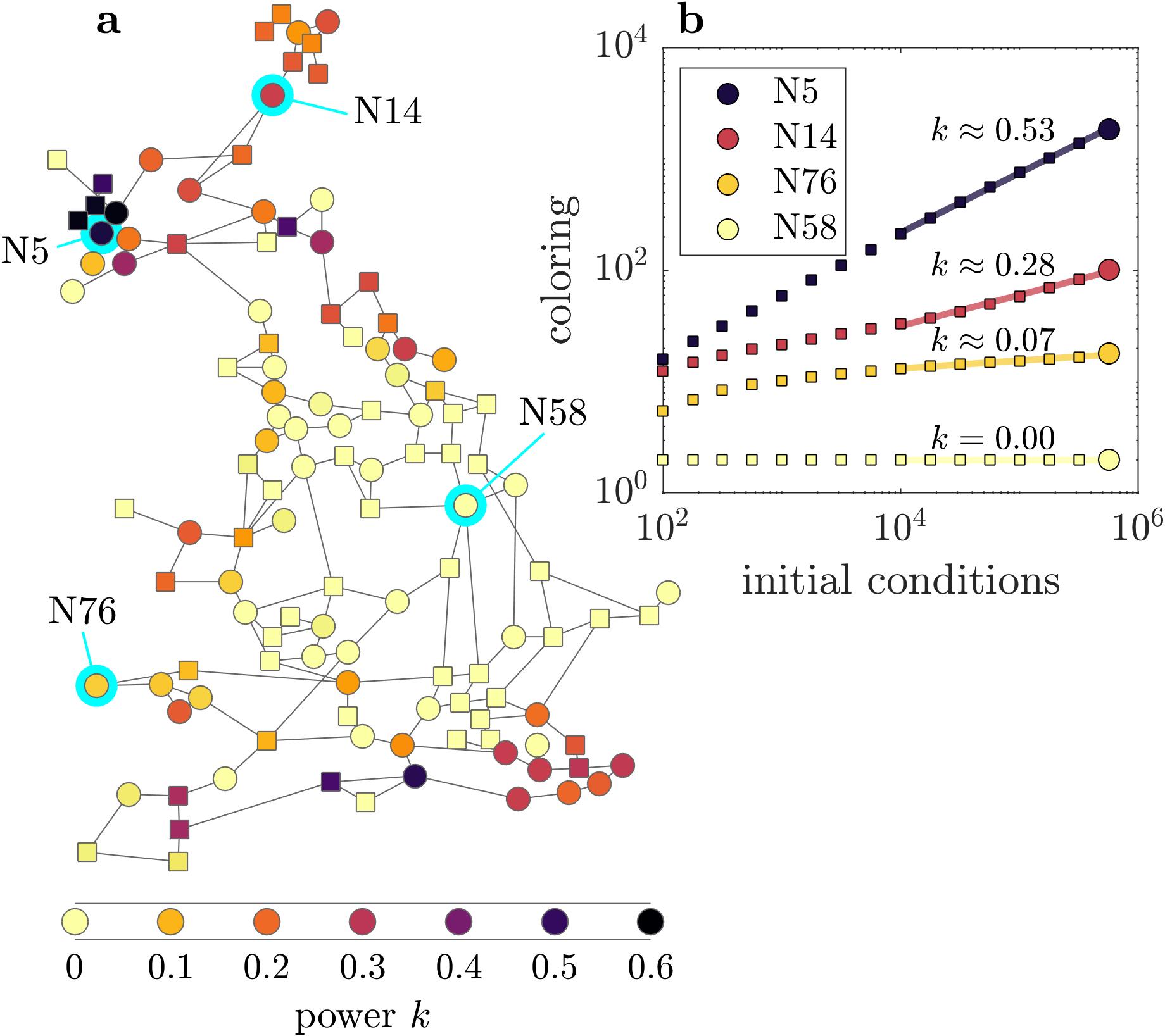}
\caption{Dependency of the coloring on the number of tested initial conditions. (a) Power $k$ which describes the increase in coloring with increasing number of initial conditions in a log-log plot. (b) Relation between the coloring and the number of tested initial conditions for four exemplary nodes. Initial conditions are drawn randomly from the complete set of initial conditions assigned to each node. Each small point is based on 100 realizations. Big points denote the coloring for the complete set of $750\times750$ initial conditions.}
\label{append_pic_a}
\end{figure}

A short reminder: In our framework, the coloring simply denotes the number of basins of attraction found in one local basin landscape. This is, however, not necessarily the same as the number of basins which are actually located in the landscape, since each landscape is only approximated by a finite set of initial conditions. In order to demonstrate the relation between the number of tested initial conditions and the coloring being detected, we determine the coloring for random subsets of the complete set of $750\times750$ initial conditions and associated basins (see Section \ref{sec_coloring}). To this end, we randomly select $N_{ini}$ initial conditions from the complete set and count the number of basins which are approached from this subset. By varying the size of the subset $N_{ini}$, we obtain a relation between the coloring and the number of initial conditions (figure \ref{append_pic_a}(b)). In consideration of the random selection, we calculate the coloring for each value of $N_{ini}$ as the mean of 100 realizations.

We find that for many nodes, the increase in coloring depending on the number of considered initial conditions is far from being saturated. Interestingly, for sufficiently large numbers of initial conditions, this relation follows approximately a power law (figure \ref{append_pic_a}(b)). The power law behavior is useful as it allows us to visualize the relation between coloring and the number of initial conditions for the whole network by approximating the power $k$ for each node (figure \ref{append_pic_a}(a)). The distribution of powers shows that the 'true' coloring could be particularly underestimated for nodes which already hold a high coloring, such as N5. Furthermore, the powers show some spatial correlation (see e.g. the neighborhood of N5 or N14). This might indicate that the local basin landscapes of certain adjacent nodes hold very similar sets of accessible attractors which are, however, embedded in complex domains of different size.

\section{Fractal basin boundaries at N5 and N14 \label{append_b}}
We have seen that the local basin landscapes of the nodes N5 and N14 contain complex regions of mixed basins or colors, respectively (figure \ref{GB_coloring}(a),(b) and figure \ref{GB_gibbs}(a),(c)). In order to verify that the contained basin boundaries are indeed fractal, we calculate the uncertainty exponent $\gamma$ \cite{grebogi1983final}. To this end, we compute the fraction $f(\epsilon)$ of initial conditions which are uncertain with respect to an initial error $\epsilon$. In this framework, an initial condition is called uncertain with respect to $\epsilon$ if it holds a different color than an associated initial condition which has a distance of $\epsilon$ (they converge to different attractors). The scaling relation between $f(\epsilon)$ and $\epsilon$ is expected to be 
\begin{equation} \label{eq_uncertain}
f(\epsilon) \sim \epsilon^{\gamma} \; ,
\end{equation} with an uncertainty exponent $\gamma<1$ denoting fractal basin boundaries.

\begin{figure}
\centering
\includegraphics[width=1.0\linewidth]{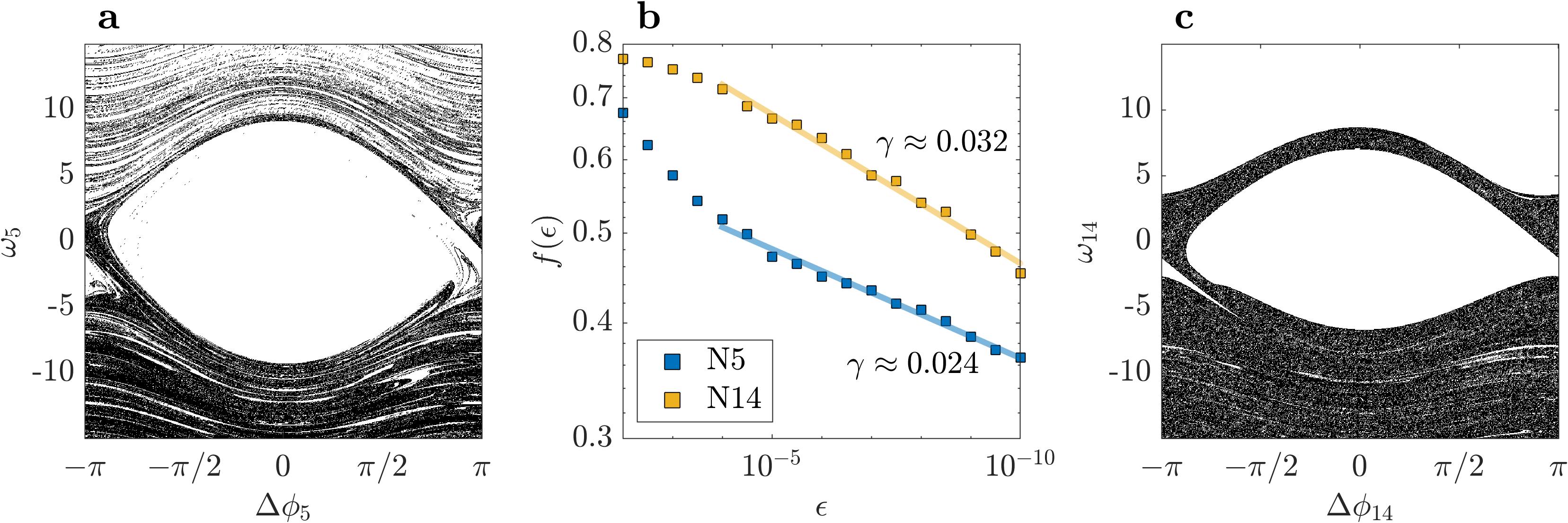}
\caption{Final state sensitivity in the mixed regions of the local basin landscapes of N5 and N14. (a,c)  Boundary boxes (areas containing basin boundaries) are colored black. (b) Relation between fraction of uncertain initial conditions and initial error $\epsilon$ within boundary boxes. The slope of the yellow and blue line gives the uncertainty exponent $\gamma$ for N5 and N14, respectively.}
\label{append_pic_b}
\end{figure}

In order to reduce the fraction of certain initial conditions, we only consider areas in which we expect to find basin boundaries. To obtain these areas, we use the set of initial conditions in the local basin landscapes (see figure \ref{GB_coloring}(a),(b)) to establish a space-filling set of boxes which are either defined as \textit{boundary boxes} or \textit{no-boundary boxes}. Each box covers the phase space between $2\times2$ adjacent initial conditions. In this sense, the initial conditions of a basin landscape constitute the corner points of the newly defined boxes. A box is defined as a boundary box if its four corner points include at least two colors and as a no-boundary box if all four converge to the same attractor. From the phase space covered by the boundary boxes (figure \ref{append_pic_b}(a),(c)), we randomly select pairs of initial conditions which have a distance of $\epsilon$ in the $\omega_i$-direction (uncertainty with respect to the frequency of node $i$). In accordance with the description in \cite{lai2011transient}, we proceed until we obtain 5000 pairs which denote an uncertain initial condition and thus obtain $f(\epsilon) \approx 5000/N_0$, with $N_0$ being the number of selected pairs. By varying the error $\epsilon$ within the interval $\epsilon \in [10^{-10},\, 10^{-2}]$, we obtain a relation between $f(\epsilon)$ and $\epsilon$ which fulfills (\ref{eq_uncertain}) for lower values of $\epsilon$ (figure \ref{append_pic_b}(b)).

\section{Trajectories on the chaotic invariant sets \label{append_c}}
\begin{figure}
\centering
\includegraphics[width=0.85\linewidth]{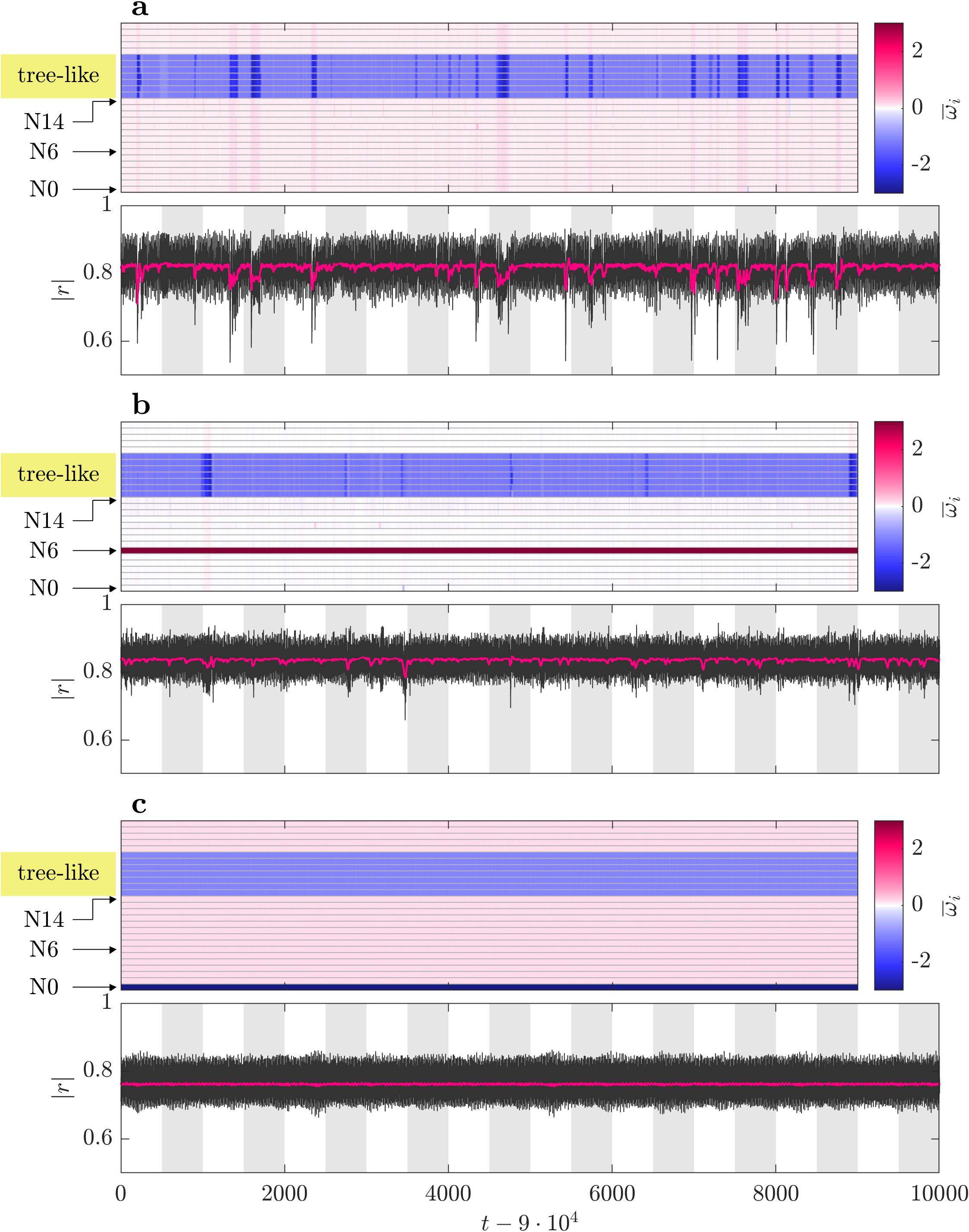}
\caption{Chaotic time series. Upper subfigures show the moving average over 25 time units of the frequencies of the 27 nodes within Scotland, including the tree-like. Lower subfigures depict the absolute value of the order parameter in dark gray and its moving average in magenta. (a) Pseudo-trajectory of the 'standard' chaotic saddle. (b) Pseudo-trajectory of the chaotic saddle in which the node N6 is solitary detached. (c) Trajectory of the chaotic attractor in which the node N0 is solitary detached.}
\label{append_pic_c}
\end{figure}

Using the stagger-and-step method \cite{sweet2001stagger}, we create pseudo-trajectories of the non-attracting invariant sets which are approached from initial conditions in the mixed/gray regions of the local basin landscapes of N14 (figure \ref{GB_saddleN14}(d)) and N5 (figure \ref{GB_saddleN5}). In figure \ref{append_pic_c}, segments of two of these pseudo-trajectories are shown. The first is the transient chaotic trajectory which we found for perturbations at N14 (figure \ref{append_pic_c}(a)). In this 'standard' case, the weakly detached tree-like cluster is established while all remaining nodes form an opposing giant cluster. The second segment shows the transient chaotic trajectory where, simultaneously to the formation of the weakly detached tree-like cluster, the node N1 is solitary detached (figure \ref{append_pic_c}(b)). Furthermore, the attracting chaotic invariant set is shown (figure \ref{append_pic_c}(c)). 

The depiction is twofold. In the upper subfigures, we demonstrate the existence of the cluster and the additional detachment of solitary nodes using the moving average of the frequencies of the single nodes. The tree-like cluster is characterized by negative mean frequencies, while the frequencies of the remaining nodes are positive. In the lower subfigures, we show the phase coherence within the whole grid based on the absolute value of the time-dependent order parameter and its moving average. The absolute value of the complex order parameter \cite{winfree2001geometry, kuramoto2003chemical} is obtained as
\begin{equation} \label{eq_order}
|r(t)| \; = \; \left| \frac{1}{N} \sum_{j=0}^{N-1} e^{\mathrm{i} \phi_j(t)} \right| \; ,
\end{equation}
where $N$ is the number of oscillators in the network and $|r|\in[0,1]$ with $|r|=1$ indicating total synchronization.

We see that if the tree-like cluster is formed and the node N0 is desychronized, the chaotic dynamics are stabilized (at least they persist for very long times; see figure \ref{GB_saddleN5}). This stabilization goes along with the disappearance of characteristic drops in the time series (figure \ref{append_pic_c}(c)) which show up in the mean frequencies of the tree-like and the phase coherence of the pseudo-trajectories (figure \ref{append_pic_c}(a),(b)). Interestingly, we find that the detachment of N1 already affects the characteristics of these drops (figure \ref{append_pic_c}(b)). The drops are lower and occur less frequently than for the 'standard' case in which no additional node is detached (figure \ref{append_pic_c}(a)). Since we also observe that the escape from the chaotic transients takes place during these drops (figure \ref{GB_saddleN14}(c)), the stabilization of the chaotic dynamics can be ascribed to the demise of these escape drops.

Finally, using the corresponding method in the JiTCODE module \cite{ansmann2018efficiently} which is based on the approach by Benettin \etal \cite{benettin1980lyapunov}, we compute the largest Lyapunov exponent $\lambda_1$ for each of the three exemplary time series (figure \ref{append_pic_a}). We find that all three exhibit a Lyapunov exponent $\lambda_1>0$: $\lambda_1 \approx 0.07$ for the pseudo-trajectory in figure \ref{append_pic_c}(a), $\lambda_1 \approx 0.06$ for figure \ref{append_pic_c}(b) and $\lambda_1 \approx 0.003$ for figure \ref{append_pic_c}(c). Accordingly, the dynamics of all three time series are chaotic, although, the chaotic signal of the dynamics which allow for the persistence of the weakly detached tree-like cluster is rather weak.

\clearpage

\section*{References}
\bibliographystyle{unsrt}
\bibliography{main.bbl}


\end{document}